\documentclass{article}

\usepackage{graphicx}
\usepackage{booktabs}
\usepackage{authblk}
\usepackage[hyphens]{url}
\usepackage[breaklinks=true]{hyperref}
\usepackage{microtype}
\usepackage{float}
\usepackage{amssymb}
\title{The Missing Adapter Layer for Research Computing}

\author{Bowen Li}
\author{Jiazhu Xie}
\author{Chelsea Wang}
\author{Alessandro Umberto D'Aloia}
\author{Ziqi Xu}
\author{Fengling Han}

\affil{RMIT University, Melbourne, Australia}

\date{March 2026}

\begin{document}

\maketitle

\begin{abstract}
Higher Degree by Research (HDR) candidates increasingly depend on
cloud-provisioned virtual machines and local GPU hardware for their
computational experiments, yet a persistent and under-addressed gap
separates \emph{having} compute resources from \emph{using} them
productively. Cloud and infrastructure teams can provision a virtual
machine in minutes, but the path from a raw VM to a reproducible,
GPU-ready research environment remains a significant barrier for
researchers who are domain experts, not systems engineers. We argue that
this gap is not a shortcoming of any particular tool but a missing
architectural layer: an \emph{adapter layer} that bridges cloud
provisioning and interactive research work.
We present a lightweight, open-source implementation of this layer, built
on k3s and Coder and already in active use in our research workspace
environment. A CI/CD pipeline connects GitHub directly to the local
cluster, carrying a research project from commit to a running, accessible
workspace in under five minutes. We then define a concrete metrics
framework for evaluating any adapter layer covering deployment latency,
environment reproducibility, onboarding friction, and resource utilisation 
and establish baselines against which improvements can be measured.
The dedicated code repository: \href{https://github.com/Aisuko/portal}{https://github.com/Aisuko/portal}
\end{abstract}

\section{Introduction}

For a Higher Degree by Research (HDR) candidate, being allocated cloud or
local GPU resources and actually running an experiment on them are two
very different milestones, often separated by days of unglamorous systems
work. A raw virtual machine arrives without the correct GPU driver stack,
without a working Python environment, and without any of the research
tooling the candidate needs. Assembling that environment requires a level
of systems knowledge that most HDR candidates domain experts in
machine learning, bioinformatics, or data science, not infrastructure
engineers simply do not have.
The result is a familiar and costly pattern: days lost at the start of
every new project configuring machines, broken environments that derail
ongoing experiments, and a disproportionate dependency on technical
support staff whose time is finite.

This friction is not caused by a shortage of compute. Universities routinely
provision cloud allocations and maintain local GPU workstations. The root
cause is structural: cloud and infrastructure tools operate at the
\emph{provisioning} layer, and they stop at the boundary of the virtual
machine. Everything that happens inside that VM driver compatibility,
environment reproducibility, self-service access, shared scheduling of
hardware is left entirely to the researcher. No managed bridge exists
between the provisioned resource and a productive working environment.

We term this the \emph{adapter layer problem}. An adapter layer is the
software infrastructure that sits between raw provisioned compute and the
interactive, reproducible research environments that HDR candidates actually
need. Without it, every researcher solves the same configuration problems
independently, every environment drifts over time, and compute resources
sit idle when they could be shared.

This paper makes three contributions. First, we formally characterise the
adapter layer problem, identifying the recurring gaps experienced by
HDR candidates across four dimensions: environment reproducibility,
self-service access, GPU resource scheduling, and vendor dependency.
Second, we describe a lightweight, open-source solution built on
k3s~\cite{k3s} and Coder~\cite{coder} and already in active production
use in our research workspace environment that implements this adapter
layer; its CI/CD pipeline connects GitHub directly to the local cluster and
deploys research projects in under five minutes, without requiring dedicated
or self-hosted CI infrastructure. Third, we propose a concrete metrics
framework for quantifying how well any adapter layer performs, and we
establish baselines for each metric that other institutions can use to
evaluate their own deployments.

Unlike managed cloud platforms such as SageMaker~\cite{sagemaker} or
Google Colab~\cite{colab}, our system supports local hardware and avoids
vendor lock-in. Unlike HPC batch schedulers~\cite{slurm,pbs}, it
prioritises interactive self-service access. Unlike notebook environments
such as JupyterHub~\cite{jupyterhub}, it provides full IDE functionality
with versioned, reproducible GPU environments. The system is deliberately
designed for small academic research teams and HDR candidate cohorts, not
large computing centres.

\section{The Adapter Layer Problem}
\label{sec:gaps}

University research computing infrastructure is typically split into two
worlds that do not communicate. The \emph{infrastructure layer} managed
by cloud teams or central IT handles provisioning: allocating virtual
machines, configuring network policies, and maintaining physical hardware.
The \emph{research layer} is what HDR candidates actually work in:
Python environments, Jupyter notebooks, GPU-accelerated training runs, and
version-controlled experiment pipelines.

Between these two worlds, nothing exists. The only interface between them
is a raw SSH connection to a freshly provisioned virtual machine;
everything above that connection the entire adapter layer is left
to the researcher to build alone.

We identify four recurring categories of gap, drawn from the daily
operation of a university research computing environment in which HDR
candidates and researchers work with provisioned compute resources.

\textbf{Environment reproducibility.}
Virtual machines provisioned from identical base images diverge rapidly
as HDR candidates independently install packages, update drivers, or
modify system configurations. There is no enforced compatibility mapping
between the host GPU driver, the CUDA runtime~\cite{nvidiaCTK}, and
machine learning framework versions. An environment that worked correctly
at the start of a project may produce different results or fail entirely
after a system update without any change to the research code itself.
This is a well-documented source of irreproducibility in computational
research~\cite{hettrick2014}.

\textbf{Onboarding friction.}
Starting a new research project or joining a new team requires an HDR
candidate to configure a complete software stack from scratch: installing
the GPU driver, matching the CUDA toolkit version to the hardware, setting
up a Python environment, and connecting a remote development interface.
Each step requires systems knowledge that most HDR candidates do not have,
creating a hard dependency on technical support staff and delaying the
start of productive research by days or weeks.

\textbf{Uncoordinated resource usage and idle capacity.}
When each HDR candidate operates an independently provisioned machine
with no shared visibility or scheduling, GPU workstations routinely sit
underutilised while other researchers wait. There is no mechanism to
reclaim idle capacity, no preemption, and no signal to identify waste.
The consequence is both inefficient use of expensive hardware and an
inequitable distribution of computing access across the research group.

\textbf{Vendor and infrastructure dependency.}
Cloud-based provisioning tools are typically tied to a single provider's
infrastructure. Local GPU workstations owned by a research group cannot
be brought into the same access model, creating a split between managed
cloud resources and unmanaged local hardware. Researchers develop
workflows that work on one but not the other, and switching between them
requires manual reconfiguration.

\begin{figure}[H]
  \centering
  \includegraphics[width=0.98\linewidth]{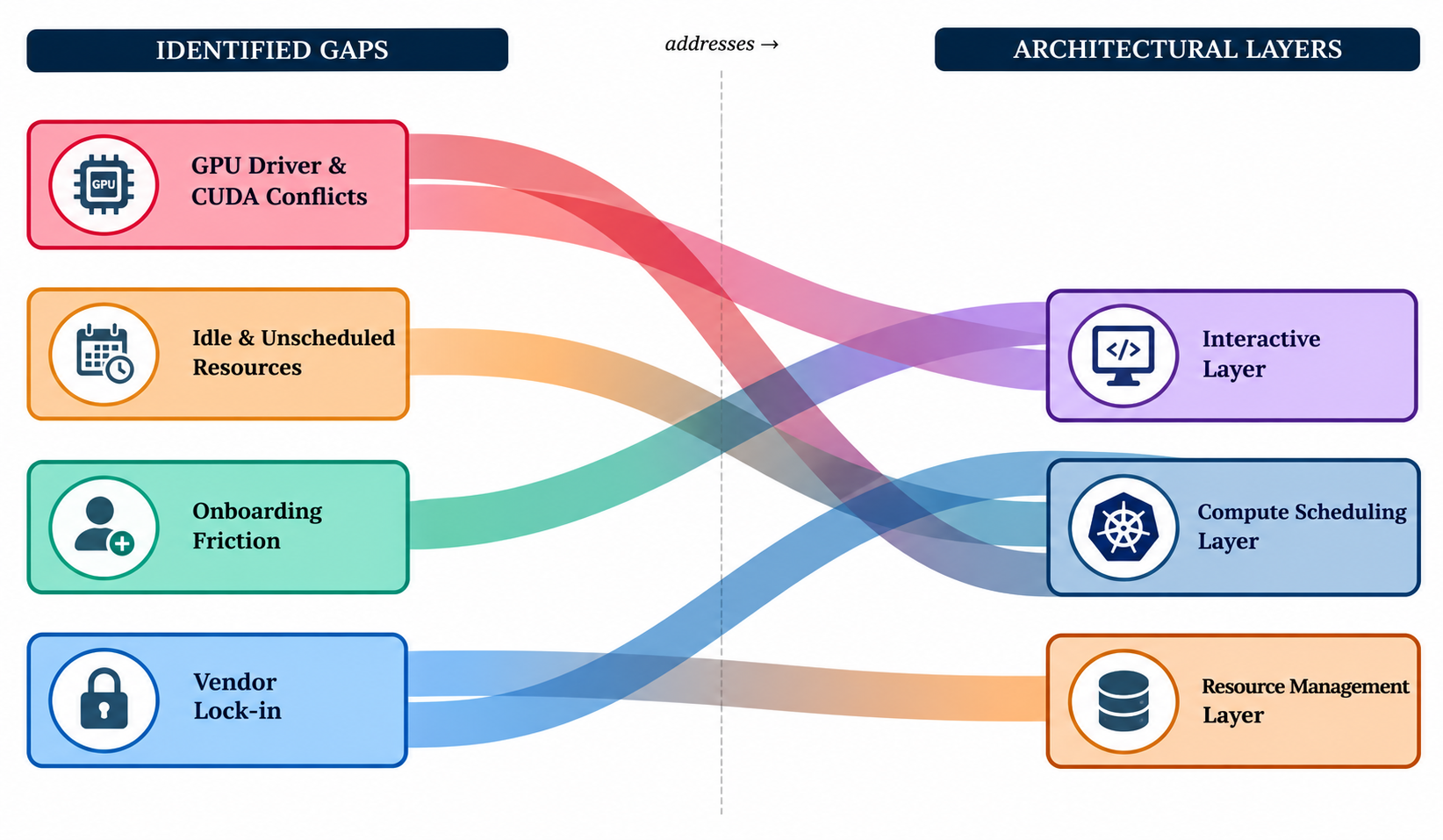}
  \caption{Gap-to-solution mapping. Each identified adapter layer gap
           maps to one or more architectural components introduced in
           Section~\ref{sec:architecture}: versioned container images,
           the k3s scheduling layer, and the Coder workspace layer.}
  \label{fig:sankey}
\end{figure}

Together, these four gaps define the adapter layer problem.
Figure~\ref{fig:sankey} shows how each one maps directly to an
architectural component in our solution. The key observation is this:
none of these gaps can be closed by improving provisioning tooling. They
are not provisioning problems at all they are symptoms of a missing
layer that sits between the provisioned resource and the researcher, and
they can only be addressed there.

\section{Architecture and Implementation}
\label{sec:architecture}

\begin{figure}[H]
  \centering
  \includegraphics[width=1.00\linewidth]{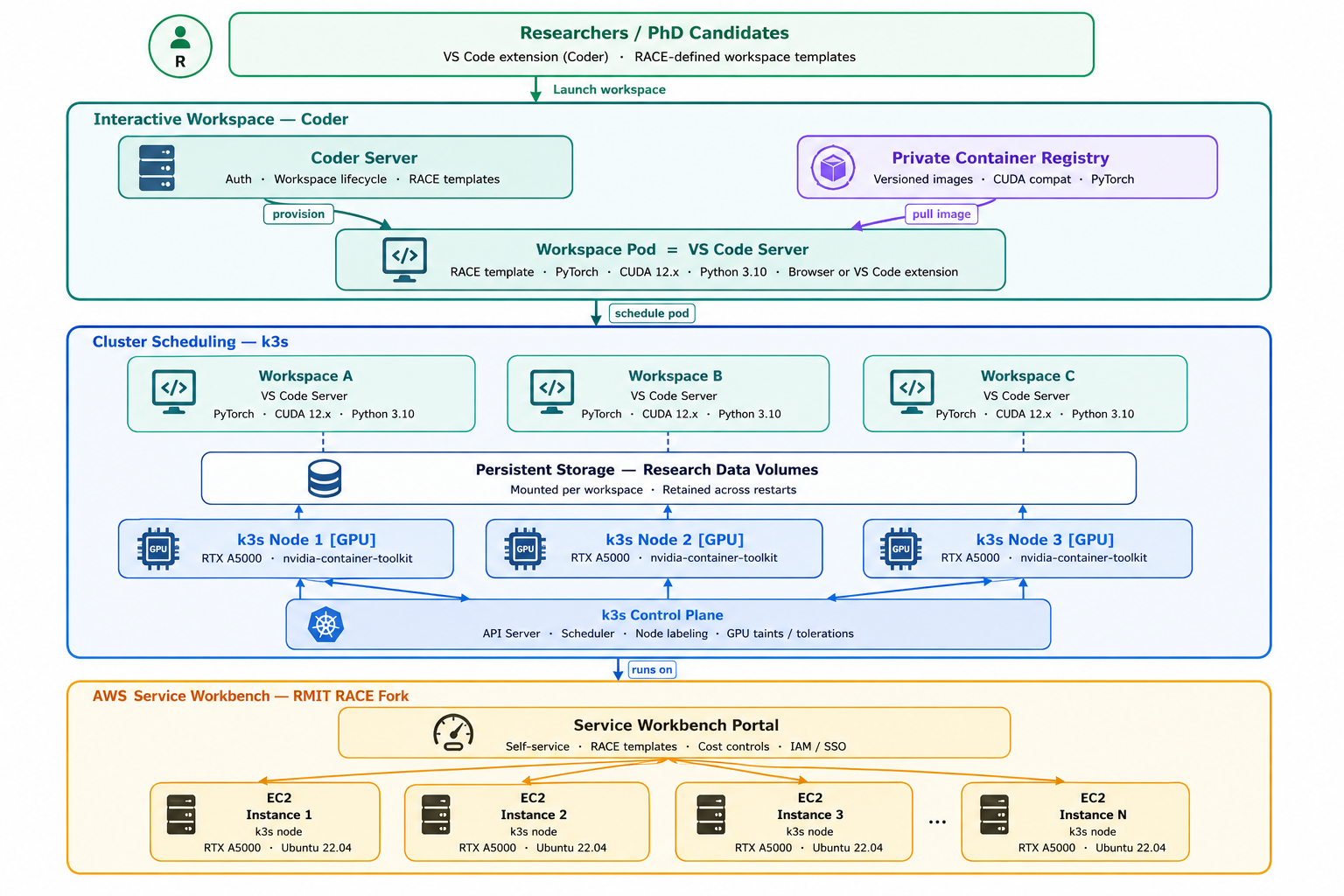}
  \caption{System architecture overview. The adapter layer (centre) sits
           between raw provisioned compute resources (EC2 instances,
           local GPU nodes) and the HDR candidate's interactive workspace.
           It comprises three components: versioned container images for
           environment reproducibility, k3s for cluster scheduling, and
           Coder for self-service workspace lifecycle management.}
  \label{fig:architecture}
\end{figure}

Having characterised the missing layer, we now describe how we build it.
Our adapter layer implementation rests on three components that address the
four gaps of Section~\ref{sec:gaps} directly, without replacing the
underlying cloud or infrastructure provisioning tools. The architecture is
illustrated in Figure~\ref{fig:architecture}.

\subsection{Cluster Layer: k3s for Lightweight Orchestration}

Local GPU workstations are pooled into a shared cluster using
k3s~\cite{k3s}, a lightweight, CNCF-certified Kubernetes
distribution~\cite{cncf} designed for resource-constrained and
operationally lean environments. Full Kubernetes was evaluated but
ruled out: a small research group without dedicated infrastructure
staff cannot absorb the operational overhead of a multi-component
control plane. k3s packages the entire control plane into a single
binary, significantly reducing both installation complexity and
ongoing maintenance burden.

Each workstation in the cluster is a GPU node equipped with an NVIDIA
RTX A5000. Nodes are registered with descriptive labels that expose
hardware capabilities to the scheduler. GPU nodes carry Kubernetes
taints~\cite{k8sTaints} so that only workloads explicitly requesting
GPU resources are scheduled onto them, preventing CPU-only workloads
from occupying scarce GPU capacity. Resource requests and limits are
enforced at the container level, ensuring that GPU memory allocation
is visible and bounded per workspace.

\subsection{Workspace Layer: Coder for Self-Service Lifecycle Management}

Researcher-facing workspaces are managed by Coder~\cite{coder}, an
open-source platform for self-hosted remote development environments.
Coder provides authentication, template-based workspace provisioning,
and full lifecycle management HDR candidates can start, stop,
rebuild, and delete their own workspaces without any staff intervention.

Workspace templates are defined and versioned centrally. Each template
specifies the container image, resource requests (CPU, memory, GPU),
and mount points for persistent storage. When an HDR candidate
provisions a workspace from a template, Coder schedules the
corresponding pod onto the k3s cluster and surfaces a
\mbox{VS~Code~Server}~\cite{vscode} interface accessible from any
browser. Candidates interact with a full VS Code environment ---
including extensions, terminals, and Git integration with no local
software installation required beyond a web browser.

The template abstraction is particularly important for HDR candidates
who are not systems engineers. A candidate selects a template by name
(e.g., \texttt{pytorch-a5000} or \texttt{tensorflow-a5000}), provisions
a workspace in approximately five minutes from a cold start, or around 20 seconds
from a warm start, and begins work
immediately in a known-good environment. The complexity of Kubernetes
scheduling, container runtimes, and driver compatibility is entirely
hidden from the researcher.

\subsection{Environment Layer: Versioned Container Images}

Reproducible software environments are enforced through a set of
curated container images maintained centrally and stored in a private
container registry. Each image encodes a complete, tested software
stack from the NVIDIA driver interface down to Python~\cite{python}
library versions and is tagged with a version identifier that
researchers and Coder templates reference explicitly.

The host nodes run NVIDIA driver version 580.126.09 with a maximum
supported CUDA version of 13.0, using \texttt{nvidia-container-toolkit}~\cite{nvidiaCTK}
and \texttt{containerd}~\cite{containerd} as the container runtime stack.
The team maintains PyTorch-focused images using CUDA runtime versions
at or below the host maximum; current supported configurations are
listed in Table~\ref{tab:compat}.

\begin{table}[h]
  \centering
  \caption{Supported container image compatibility matrix.}
  \label{tab:compat}
  \begin{tabular}{llll}
    \toprule
    \textbf{Image Tag} & \textbf{CUDA Runtime} & \textbf{Framework}
      & \textbf{Python} \\
    \midrule
    \texttt{pytorch-2x-cu121}   & 12.1 & PyTorch 2.10 & 3.10 \\
    \texttt{pytorch-2x-cu124}   & 12.4 & PyTorch 2.10 & 3.10 \\
    \texttt{pytorch-2x-cu130}   & 13.0 & PyTorch 2.10 & 3.10 \\
    \bottomrule
  \end{tabular}
\end{table}

When a new GPU driver is deployed to host nodes, the team updates and
publishes a new set of images against the updated compatibility matrix.
Existing image tags are preserved so that HDR candidates can continue
to use prior environments without disruption.

\subsection{CI/CD Pipeline: Project Deployment in Under Five Minutes}
\label{sec:cicd}

A key operational claim of this adapter layer is that research projects
can be deployed to the local cluster in under five minutes via our CI/CD
pipeline. This pipeline connects GitHub directly to the local cluster,
automating the full path from a project repository
to a running, accessible workspace: validating code quality, building
and pushing the container image to the registry, and deploying the
application to the k3s cluster via Helm all triggered by a single
\texttt{git push}.

The pipeline is structured as three sequential stages:

\begin{enumerate}
  \item \textbf{Validate} ($\sim$60--90 seconds): Code quality checks
        including linting, type checking, and Helm chart validation run
        in parallel where possible. This stage fails fast on errors before
        any image build is attempted.
  \item \textbf{Build and push} ($\sim$60--120 seconds): The container
        image is built using Docker Buildx with GitHub Actions layer
        caching (\texttt{cache-from: type=gha}). Cache hits reduce this
        stage significantly; cache misses on large dependency layers
        dominate the upper bound of total pipeline time.
  \item \textbf{Deploy to cluster} ($\sim$60--90 seconds): The pipeline
        connects to the local cluster, syncs
        Kubernetes secrets, and executes a \texttt{helm upgrade --install}
        against the target namespace. The deployment is complete when
        the pod reaches \texttt{Running} state.
\end{enumerate}

To validate the under-five-minute claim empirically, we measured end-to-end
pipeline duration across ten consecutive runs for each of three production
projects currently deployed on the cluster, spanning different application
types and pipeline structures. Results are shown in
Table~\ref{tab:cicd_times}.

\begin{table}[H]
  \centering
  \caption{CI/CD pipeline end-to-end deployment times across three
           production projects, measured over ten consecutive runs each
           on free-tier GitHub Actions runners. All runs deploy to the
           local k3s cluster via Helm. Build cache refers to GitHub
           Actions layer caching (\texttt{type=gha}).}
  \label{tab:cicd_times}
  \begin{tabular}{p{2.2cm} p{3.2cm} p{1.4cm} p{1.6cm} p{1.6cm}}
    \toprule
    \textbf{Project} & \textbf{Pipeline stages} & \textbf{Build cache} &
    \textbf{Time range} & \textbf{All $<$5 min} \\
    \midrule
    Project A
      & lint $\to$ build $\to$ push $\to$ Helm deploy
      & GHA cache
      & 3m\,21s -- 3m\,40s
      & \checkmark \\[4pt]
    Project B
      & quality $\to$ build $\to$ push $\to$ Helm deploy
      & GHA cache
      & 2m\,51s -- 3m\,51s
      & \checkmark \\[4pt]
    Project C
      & lint $\to$ test $\to$ build $\to$ push $\to$ CRD deploy
      & GHA cache
      & 4m\,00s -- 5m\,00s
      & \checkmark \\
    \bottomrule
  \end{tabular}
\end{table}

All three projects a CPU web application, a GPU-backed web application,
and a cloud-native Kubernetes operator with custom resource definitions ---
complete the full validate, build, and cluster deployment cycle within five
minutes on free-tier GitHub Actions runners. Project~C represents the most
complex pipeline (five stages including CRD deployment) and still falls
within the bound. These results demonstrate that the five-minute deployment
claim holds across heterogeneous project types without requiring dedicated
or self-hosted CI infrastructure.

\section{Metrics Framework and Baselines}
\label{sec:metrics}

Characterising the adapter layer problem requires not just an architecture
but a way to measure whether that architecture is solving the problem.
We define a metrics framework covering four dimensions that directly
correspond to the four gaps identified in Section~\ref{sec:gaps}, and
we establish concrete baselines against which the system's performance
can be evaluated and compared.

\subsection{Metric 1: Workspace Deployment Latency}

\textbf{Definition.} The elapsed time from an HDR candidate initiating a
workspace request (clicking ``Create Workspace'' in the Coder interface, or
triggering the CI/CD pipeline) to the moment the \mbox{VS~Code~Server}
environment is accessible and ready for use.

\textbf{Measurement approach.} We distinguish three conditions:
\begin{itemize}
  \item \emph{Cold start}: container image not present on any node; full
        image pull required before initialisation.
  \item \emph{Warm start}: image already cached on the target node from
        a prior pull; only container initialisation and VS~Code~Server
        startup time.
  \item \emph{CI/CD pipeline deployment}: new project added via the
        automated pipeline, including template registration.
\end{itemize}

\textbf{Baseline.} The baseline is end-to-end cloud VM provisioning, which
represents the path an HDR candidate would take without the adapter layer:
request a VM, wait for it to boot, then manually configure the software
environment. Cloud VM provisioning typically takes 10--20 minutes to reach
a running instance, after which manual environment setup adds a further
30--90 minutes for a new project. Table~\ref{tab:latency} summarises
measured deployment times against this baseline.

\begin{table}[h]
  \centering
  \caption{Workspace deployment latency across startup conditions.
           Baseline is end-to-end cloud VM provisioning to a usable
           research environment (including manual setup).}
  \label{tab:latency}
  \begin{tabular}{lll}
    \toprule
    \textbf{Method} & \textbf{Deployment Time} & \textbf{Notes} \\
    \midrule
    Cloud VM (baseline)         & $\sim$10--20 min  & VM boot only \\
    Adapter layer, cold start   & $\sim$5 min        & Image pull +
                                                       container init \\
    Adapter layer, warm start   & $\sim$20 s         & Image cached on node \\
    CI/CD pipeline, warm cache  & $<$5 min           & Full project deploy \\
    \bottomrule
  \end{tabular}
\end{table}

These figures confirm that for researchers returning to an active project,
warm-start access is effectively immediate.

\subsection{Metric 2: Environment Reproducibility Rate}

\textbf{Definition.} The proportion of workspace starts in which the
researcher's environment matches the expected software stack correct
GPU driver binding, correct CUDA runtime version, and correct ML
framework version without any manual intervention.

\textbf{Measurement approach.} At each workspace start, an automated
health check runs inside the container and records: (i) whether
\texttt{nvidia-smi} reports the expected driver version, (ii) whether
the CUDA runtime version matches the image tag, and (iii) whether the
primary ML framework (PyTorch or TensorFlow) imports without error.
A workspace is counted as \emph{reproducible} if all three checks pass.

\textbf{Baseline.} Without the adapter layer, environment reproducibility
is not enforced. Each HDR candidate manages their own environment
independently, and environment drift is ubiquitous. Surveys of research
software practice~\cite{hettrick2014} consistently find that environment
configuration problems are among the most common sources of wasted time
in computational research. The baseline reproducibility rate without a
managed layer is therefore indeterminate: no systematic measurement is
possible without the measurement infrastructure the adapter layer itself
provides. This indeterminacy is itself evidence of the gap the absence
of enforcement makes the problem unmeasurable and therefore unmanageable.

\textbf{Target.} A well-functioning adapter layer should achieve a
reproducibility rate of $\geq$99\% across workspace starts. Any failure
indicates either a broken image or a driver/toolkit mismatch on the host
node, both of which are actionable maintenance signals.

\subsection{Metric 3: Onboarding Time to First Experiment}

\textbf{Definition.} The elapsed time from an HDR candidate's first
interaction with the system (account creation or first workspace
provisioning) to their first successful execution of research code ---
defined as a training script or data processing pipeline completing
without environment-related errors.

\textbf{Measurement approach.} Onboarding time is recorded per researcher
at the workspace level: the timestamp of first workspace creation and
the timestamp of the first successful job completion (exit code 0) in
that workspace. The difference is the onboarding time. Researchers who
require support tickets or staff intervention are flagged separately as
\emph{assisted onboardings}.

\textbf{Baseline.} Without the adapter layer, onboarding a new HDR
candidate to a functional GPU environment typically involves:
(i) requesting VM access from cloud or IT teams (1--3 business days),
(ii) manual GPU driver and CUDA installation (2--4 hours, often
requiring support staff), and (iii) Python environment configuration
(1--2 hours). A realistic baseline for time to first experiment is
therefore 1--3 business days, with significant variance driven by
staff availability and the candidate's technical background.

\subsection{Metric 4: GPU Utilisation and Idle Capacity}

\textbf{Definition.} The fraction of available GPU-hours across the
cluster that are actively utilised by running workloads, measured over
a rolling 7-day window. The complement idle GPU-hours quantifies
wasted capacity.

\textbf{Measurement approach.} GPU utilisation is sampled at one-minute
intervals from each node using the NVIDIA Management Library (NVML)
via \texttt{dcgm-exporter}~\cite{nvidiaCTK} integrated with a Prometheus
and Grafana monitoring stack. Utilisation is recorded as the percentage
of GPU compute units active during each sample interval. Idle capacity
is defined as nodes with $<$5\% GPU utilisation for a sustained
30-minute window, indicating no active workload.

\textbf{Baseline.} In a model where each HDR candidate occupies a
dedicated, unmanaged VM, GPU utilisation is determined entirely by
individual usage patterns. Typical academic workloads are highly
bursty: intensive during active training runs and near-zero between
experiments. Without scheduling or reclamation, the expected average
utilisation across a fleet of individually allocated machines is
typically below 30\% in academic settings. The adapter layer's shared
scheduling model makes idle capacity visible and actionable.

\subsection{Metrics Summary}

Table~\ref{tab:metrics_summary} consolidates the four metrics, their
measurement methods, baselines, and targets. This framework is designed
to be replicable: any institution deploying a comparable adapter layer
can apply the same measurements to evaluate their own deployment.

\begin{table}[H]
  \centering
  \caption{Adapter layer metrics framework: definitions, baselines, and targets.}
  \label{tab:metrics_summary}
  \begin{tabular}{p{2.8cm} p{3.0cm} p{3.0cm} p{2.8cm}}
    \toprule
    \textbf{Metric} & \textbf{Baseline (no adapter layer)} &
    \textbf{Target} & \textbf{Measurement} \\
    \midrule
    Deployment latency
      & 10--20 min (VM boot only)
      & $<$20 s (warm), $<$5 min (CI/CD)
      & Timestamp: request $\to$ VS~Code ready \\[4pt]
    Reproducibility rate
      & Indeterminate (no enforcement)
      & $\geq$99\% of workspace starts
      & Automated health check at start \\[4pt]
    Onboarding time
      & 1--3 business days
      & To be established
      & First workspace $\to$ first successful run \\[4pt]
    GPU utilisation
      & $<$30\% (dedicated, unmanaged)
      & To be established
      & NVML / dcgm-exporter, 1-min intervals \\
    \bottomrule
  \end{tabular}
\end{table}

\section{Related Work}
\label{sec:related}

\textbf{Managed cloud research platforms.}
Platforms such as AWS SageMaker~\cite{sagemaker}, Google Colab~\cite{colab},
and Microsoft Azure Machine Learning~\cite{azureml} provide managed
environments for data science and machine learning workflows. These
platforms offer convenient access to cloud compute and pre-configured
runtimes, but are tightly coupled to their respective cloud providers,
with no mechanism to incorporate local or on-premises hardware.
Their cost and governance models are not designed for the shared,
team-managed environment typical of university research groups.
Crucially, these platforms do not expose the adapter layer as a
composable or portable component: the abstraction is available only
inside the provider's ecosystem.

\textbf{HPC schedulers and cluster portals.}
Traditional high-performance computing infrastructure relies on batch
schedulers such as SLURM~\cite{slurm} and PBS~\cite{pbs}. These systems
were designed for large-scale batch job queues~\cite{slurm,pbs}, and as
a result present significant administrative overhead and a steep learning
curve for small research teams whose primary workflow is interactive
development rather than job submission~\cite{hettrick2014}. Open
OnDemand~\cite{openondemand} provides a web portal layer over HPC clusters
but still requires substantial institutional infrastructure to operate
and does not address the environment reproducibility problem.

\textbf{Notebook and cloud IDE environments.}
JupyterHub~\cite{jupyterhub} is widely deployed in academic settings
and provides multi-user notebook environments that lower the barrier
to interactive computing. However, JupyterHub does not natively
provide full IDE functionality, workspace lifecycle management, or
the versioned environment control required for reproducible GPU
workloads~\cite{jupyterhub}. GitHub Codespaces~\cite{codespaces} offers
a cloud-hosted VS Code experience but is designed exclusively for
GitHub's cloud infrastructure and cannot be deployed on local
hardware~\cite{codespaces}. Neither constitutes a general-purpose
adapter layer: both assume that environment configuration is already
solved.

\textbf{Container orchestration for research computing.}
Kubeflow~\cite{kubeflow} and similar platforms apply Kubernetes
orchestration to machine learning pipelines. These systems are designed
for structured ML pipelines rather than general interactive compute~\cite{kubeflow},
and their operational footprint spanning multiple microservices
and custom resource definitions is typically substantial, requiring
dedicated DevOps expertise to install and maintain~\cite{bisong2019}.
The complexity is the opposite of what an adapter layer for HDR
candidates should present.

\textbf{Positioning.}
Our solution is, to our knowledge, the first lightweight, open-source,
and vendor-neutral implementation designed specifically for HDR candidates
in small university research teams that explicitly frames this as an
adapter layer problem. The combination of k3s for cluster scheduling,
Coder for self-service workspace management, and versioned container
images for reproducibility addresses all four identified gaps within a
system that any research group with a single capable person can install
and operate.

\section{Discussion and Conclusion}
\label{sec:conclusion}

\subsection{Discussion}

The adapter layer framing matters because it makes the problem
tractable. Rather than arguing that cloud provisioning tools should be
extended to manage environments, or that HPC schedulers should be
simplified for interactive use, we isolate a distinct architectural
concern and address it with a purpose-built layer. The four gaps of
Section~\ref{sec:gaps} are not provisioning failures that better
provisioning could fix; naming them collectively as the adapter layer is
what turns a diffuse set of researcher frustrations into a single,
addressable design target.

The deployment in our research workspace environment demonstrates that
this layer can be built and operated with realistic resources. The system
uses only open-source components, runs on hardware already available in
the research group, and does not require a dedicated infrastructure team.
The CI/CD pipeline that connects GitHub directly to the local cluster,
delivering research project deployments in under five minutes, is a
concrete operational outcome, not a theoretical claim.

The metrics framework in Section~\ref{sec:metrics} is a contribution
in its own right. Characterising the adapter layer problem quantitatively
 with baselines drawn from the without-adapter-layer state and
targets that reflect what is achievable provides a vocabulary for
evaluating and comparing solutions. Institutions that deploy similar
systems can apply the same framework to assess whether their own adapter
layer is working.

\textbf{Limitations.}
The initial cluster and Coder deployment requires a period of
infrastructure work best suited to teams with at least one technically
experienced member. Integration with institutional SSO systems requires
coordination with university IT and represents a one-time setup cost.
The current system also lacks fine-grained quota enforcement across
researchers, which is the primary area for near-term improvement as the
cluster grows.

\subsection{Future Work}

Several directions would strengthen the system. Improved scheduling
policies including bin-packing and preemption for lower-priority
workloads would increase GPU utilisation under contention.
Stronger governance tooling, such as per-researcher GPU quotas and
automated idle workspace reclamation, would reduce waste without
requiring manual monitoring. Longer term, federating the adapter layer
across multiple research labs within the university would allow hardware
resources to be shared more broadly while preserving team-level access
controls.

The metrics framework should also be extended with longitudinal data
to assess whether improvements in deployment latency and onboarding
time translate into measurable increases in research output paper
submissions, experiment iteration rate, or time-to-result for common
HDR milestones.

\subsection{Conclusion}

We have characterised and addressed a missing adapter layer in university
research computing infrastructure. HDR candidates need more than
provisioned compute they need a bridge from raw resources to
reproducible, self-service, GPU-ready working environments. This bridge
does not currently exist in standard provisioning tooling.

Our open-source implementation, built on k3s~\cite{k3s} and
Coder~\cite{coder} and already in active use in our research workspace
environment, demonstrates that this adapter layer is lightweight,
installable, and operational with a CI/CD pipeline that carries a
project from a GitHub commit to a running workspace on the local cluster
in under five minutes. We have additionally proposed a concrete metrics
framework deployment latency, environment reproducibility, onboarding
time, and GPU utilisation with baselines that enable rigorous
evaluation and replication by other institutions. We hope this work serves
as both a practical reference and a call to recognise the adapter layer as
a first-class concern in research computing infrastructure one that
deserves the same deliberate design attention as the provisioning layer
beneath it.

\bibliographystyle{unsrt}
\bibliography{main}

@misc{k3s,
  author       = {{Rancher Labs}},
  title        = {k3s: Lightweight {Kubernetes}},
  year         = {2024},
  url          = {https://k3s.io},
  note         = {Accessed: March 2026}
}

@misc{coder,
  author       = {{Coder Technologies}},
  title        = {Coder: Self-Hosted Remote Development Environments},
  year         = {2024},
  url          = {https://coder.com},
  note         = {Accessed: March 2026}
}

@misc{vscode,
  author       = {{Microsoft}},
  title        = {Visual Studio Code},
  year         = {2024},
  url          = {https://code.visualstudio.com},
  note         = {Accessed: March 2026}
}

@misc{jupyterhub,
  author       = {{Project Jupyter}},
  title        = {{JupyterHub}},
  year         = {2024},
  url          = {https://jupyter.org/hub},
  note         = {Accessed: March 2026}
}

@misc{nvidiaCTK,
  author       = {{NVIDIA Corporation}},
  title        = {{NVIDIA} Container Toolkit},
  year         = {2024},
  url          = {https://docs.nvidia.com/datacenter/cloud-native/container-toolkit/overview.html},
  note         = {Accessed: March 2026}
}

@misc{containerd,
  author       = {{Cloud Native Computing Foundation}},
  title        = {containerd: An Industry-Standard Container Runtime},
  year         = {2024},
  url          = {https://containerd.io},
  note         = {Accessed: March 2026}
}

@misc{cncf,
  author       = {{Cloud Native Computing Foundation}},
  title        = {Cloud Native Computing Foundation},
  year         = {2024},
  url          = {https://www.cncf.io},
  note         = {Accessed: March 2026}
}

@misc{k8sTaints,
  author       = {{Kubernetes Authors}},
  title        = {Taints and Tolerations},
  year         = {2024},
  url          = {https://kubernetes.io/docs/concepts/scheduling-eviction/taint-and-toleration/},
  note         = {Accessed: March 2026}
}

@misc{python,
  author       = {Van Rossum, Guido and Drake, Fred L.},
  title        = {Python 3 Reference Manual},
  year         = {2009},
  publisher    = {CreateSpace},
  address      = {Scotts Valley, CA},
  url          = {https://www.python.org}
}

@misc{sagemaker,
  author       = {{Amazon Web Services}},
  title        = {{Amazon SageMaker}},
  year         = {2024},
  url          = {https://aws.amazon.com/sagemaker/},
  note         = {Accessed: March 2026}
}

@misc{colab,
  author       = {{Google}},
  title        = {Google Colaboratory},
  year         = {2024},
  url          = {https://colab.research.google.com},
  note         = {Accessed: March 2026}
}

@misc{azureml,
  author       = {{Microsoft}},
  title        = {Azure Machine Learning},
  year         = {2024},
  url          = {https://azure.microsoft.com/en-us/products/machine-learning},
  note         = {Accessed: March 2026}
}

@inproceedings{slurm,
  author       = {Yoo, Andy B. and Jette, Morris A. and Grondona, Mark},
  title        = {{SLURM}: Simple {Linux} Utility for Resource Management},
  booktitle    = {Job Scheduling Strategies for Parallel Processing (JSSPP)},
  year         = {2003},
  publisher    = {Springer},
  address      = {Berlin, Heidelberg},
  pages        = {44--60},
  doi          = {10.1007/10968987_3}
}

@misc{pbs,
  author       = {{Altair Engineering}},
  title        = {{OpenPBS}: Open Source High Performance Computing Workload Manager},
  year         = {2024},
  url          = {https://www.openpbs.org},
  note         = {Accessed: March 2026}
}

@article{openondemand,
  author       = {Hudak, Dave and Johnson, Doug and Chalker, Alan and
                  Nicklas, Jeremy and Franz, Eric and Dockendorf, Trey and
                  McMichael, Brian L.},
  title        = {Open {OnDemand}: A Web-Based Client Portal for {HPC} Centers},
  journal      = {Journal of Open Source Software},
  year         = {2018},
  volume       = {3},
  number       = {25},
  pages        = {622},
  doi          = {10.21105/joss.00622}
}

@misc{codespaces,
  author       = {{GitHub}},
  title        = {{GitHub Codespaces}},
  year         = {2024},
  url          = {https://github.com/features/codespaces},
  note         = {Accessed: March 2026}
}

@misc{kubeflow,
  author       = {{Kubeflow Authors}},
  title        = {Kubeflow: Machine Learning Toolkit for {Kubernetes}},
  year         = {2024},
  url          = {https://www.kubeflow.org},
  note         = {Accessed: March 2026}
}

@misc{hettrick2014,
  author       = {Hettrick, Simon and Antonioletti, Mario and Carr, Les and
                  {Chue Hong}, Neil and Crouch, Stephen and {De Roure}, David
                  and Emsley, Iain and Goble, Carole and Hay, Alexander and
                  Inupakutika, Devasena and Jackson, Mike and
                  Nenadic, Aleksandra and Parkinson, Tim and
                  Parsons, Mark I. and Pawlik, Aleksandra and Peru, Giacomo
                  and Proeme, Arno and Robinson, John and Sufi, Shoaib},
  title        = {{UK} Research Software Survey 2014},
  year         = {2014},
  publisher    = {Zenodo},
  doi          = {10.5281/zenodo.14809},
  url          = {https://doi.org/10.5281/zenodo.14809}
}

@incollection{bisong2019,
  author       = {Bisong, Ekaba},
  title        = {Kubeflow and Kubeflow Pipelines},
  booktitle    = {Building Machine Learning and Deep Learning Models on
                  {Google Cloud Platform}},
  publisher    = {Apress},
  address      = {Berkeley, CA},
  year         = {2019},
  pages        = {671--685},
  doi          = {10.1007/978-1-4842-4470-8_46}
}

\end{document}